\documentclass[aps,prd,preprintnumbers,showpacs,showkeys,nofootinbib,
superscriptaddress,fleqn,floatfix,tightenlines, 10pt]{revtex4-1}
\usepackage{amsmath,amsfonts,amssymb,amscd,amsxtra,amsthm}
\usepackage{graphicx}  
\usepackage[dvipsnames]{xcolor} 
\newcommand{\ket}{\rangle}
\newcommand{\bra}{\langle}
\begin{document}  
\preprint{INHA-NTG-02/2019}
\title{Strange and Charmed Baryon Productions with an Instanton Interaction}
\author{Sang-In Shim}
\email[E-mail: ]{shimsang@rcnp.osaka-u.ac.jp}
\affiliation{Research Center for Nuclear Physics (RCNP),
Osaka University, Ibaraki, Osaka, 567-0047, Japan}
\author{Atsushi Hosaka}
\email[E-mail: ]{hosaka@rcnp.osaka-u.ac.jp}
\affiliation{Research Center for Nuclear Physics (RCNP),
Osaka University, Ibaraki, Osaka, 567-0047, Japan}
\author{Hyun-Chul Kim}
\email[E-mail: ]{hchkim@inha.ac.kr}
\affiliation{Department of Physics, Inha University, Incheon 22212,
Republic of Korea}
\affiliation{Advanced Science Research Center, Japan Atomic Energy
  Agency, Shirakata, Tokai, Ibaraki, 319-1195, Japan}
\affiliation{School of Physics, Korea Institute for Advanced Study 
  (KIAS), Seoul 02455, Republic of Korea}
\date{\today}
\begin{abstract}
We study strange and charmed baryon productions, $\pi + p
  \rightarrow K(D) + Y(Yc)$, where $Y(Yc)$ is a strange(charmed)
  baryon of a ground or an excited state. We propose a new production
  mechanism where two quarks in the baryon participate in the
  reaction (two-quark process), which enables excitation of both
  $\lambda$- and $\rho$-modes of the baryons. To deal with the
  two-quark process, we consider the 't Hooft six-quark interaction
  from the instanton model. We study production rates in relation with
  the structure of charmed baryons for the forward angle scattering. 
\end{abstract}
\maketitle

\section{Introduction}
Recently, as new hadrons including heavy quarks such as charm and
bottom have been observed continuously, experimental and theoretical
researches on them have been actively
conducted~\cite{Hosaka:2016pey}. It is necessary to fix their
properties such as spin-parity, structure, and reaction mechanism of
hadrons. In this context, a heavy baryon can be a good testing
ground for a research. In a baryon containing one heavy quark, two
types of excited states from the heavy quark-diquark description, the
so-called $\lambda$- and $\rho$-modes, are distinguished, providing
unique features for heavy baryon structure together with heavy quark
symmetry~\cite{Yoshida:2015tia}. 

Based on the above background, an experimental plan for charmed baryon
productions, $\pi^-+p \rightarrow D^*+Y_c$, has been made at
J-PARC~\cite{e50} and the theoretical studies on this experiment have
been carried out~\cite{Kim:2014qha}. However, in the previous
theoretical work, only the $\lambda$-modes are discussed among the two
kinds of excited states of heavy baryons, a further theoretical study is
required to investigate both $\lambda$- and $\rho$-mode excitations.  

In the present work we propose a reaction mechanism for heavy baryon
productions, $\pi^-+p \rightarrow K^0 + Y$ or $D^- +Y_c$, which can
excite both $\lambda$- and $\rho$-modes. We calculate matrix elements
for the reaction by using a non-relativistic constituent quark model
with a three-quark interaction derived from the QCD instanton
model. We also calculate the production rates of various heavy
baryons. 

\section{Formalism}
\subsection{Two-quark process}
\begin{figure}[htp]
\centering
\includegraphics[width = 10cm]{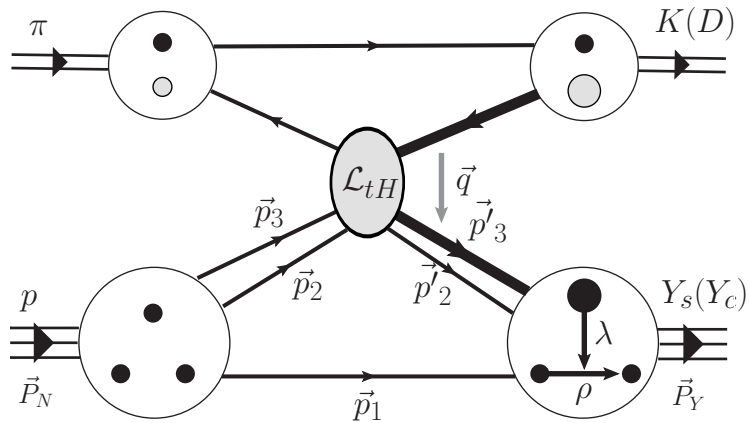}
\caption{two-quark processes for heavy baryon productions. Two quarks
  in the target baryon participate in the reaction and the final state
  heavy baryon can be excited to the $\lambda$-modes and to the
  $\rho$-modes.} 
\label{fig:1}
\end{figure}

In Fig.~\ref{fig:1}, a quark line representation for two-quark process
is given. In the two-quark process, an antiquark in the pion interacts
with two quarks in the proton. The two-quark process excites both
$\lambda$- and $\rho$-modes. Here, $\lambda$-modes are excitations of
relative motion between a diquark and a heavy quark and $\rho$-modes
are excitations of diquark itself in the heavy baryon. One can also
consider one-quark process in which one quark in the baryon interacts
with an antiquark in the pion and possible excitations are only
$\lambda$-modes~\cite{Kim:2014qha}. 

In Fig.~\ref{fig:1}, various momentum fractions carried by various
quarks are shown; the momenta of the initial and the final state
baryons consist of the momenta of the three quarks inside of the
baryons, $\vec{P}_N=\vec{p}_1+\vec{p}_2+\vec{p}_3$,
$\vec{P}_Y=\vec{p}_1+\vec{p'}_2+\vec{p'}_3$, where $\vec{p}_i$ and
$\vec{p'}_i=\vec{p}_i+\vec{q}_i$ ($i=1,2,3$) are the quark momenta
inside of the baryons and $\vec{q}_i$ is the transferred momentum from
the initial pion to i-th quark in the heavy baryon. In the two quark
process the momentum transfer $\vec{q}$ is shared by two quarks (2,3)
such that $\vec{q} = \vec{P}_Y-\vec{P}_p = \vec{q}_2 + \vec{q}_3$ is
the transferred momentum from the pion to the heavy baryon. 

\subsection{Three-quark interaction}

For the description of two-quark processes we need a suitable
interaction where three quarks participate in. 
Here we employ a point-like interaction of three quarks, the 't Hooft
interaction, as inspired by the instanton dynamics of QCD.   
The interaction works well for low-energy hadron properties including
up, down and strange quarks. However, a naive extension to the charm
sector may not be available, while some discussions have been
made~\cite{Chernyshev:1994-1995}. Thus, the present study is most
likely to be applied to the strange sector, though some features may
be  discussed for the charm productions. Therefore, results will be
shown both for strangeness and charm productions. 

The use of the zero-range three-body interaction has a virtue  
that actual computations become simple. For a more realistic
description, finite-range nature of the interaction may be included in 
terms of form factors.  

The relevant 'tHooft interaction is given by~\cite{tHooft:1976}. 
\begin{align}
\mathcal{L}_{tH}
= c \,\mathrm{det}[\bar{q}_i (1+\gamma_5)q_j] + H.c. 
= c \left| \begin{array}{ccc}
\bar{u}(1+\gamma_5)u & \bar{u}(1+\gamma_5)d & \bar{u}(1+\gamma_5)s \\
\bar{d}(1+\gamma_5)u & \bar{d}(1+\gamma_5)d & \bar{d}(1+\gamma_5)s \\
\bar{s}(1+\gamma_5)u & \bar{s}(1+\gamma_5)d & \bar{s}(1+\gamma_5)s
 \end{array} \right| + H.c.,
 \label{eq:tHooft}
\end{align}
where $c$ is a coupling constant, which, however, is not important in
the present study, since it is difficult to determine the absolute
values of the reaction cross sections. Thus, we focus here only on
relative production rates. 

\section{Matrix elements and production rates}
In a non-relativistic quark model, baryon wave functions can be
written as a product of the plane wave for the center-of-mass motion
and bound state wave functions in internal coordinates, Jacobi
coordinates $\lambda$ and $\rho$. Then, the transition amplitudes for
the reaction $\pi^- p \to M Y$ can be written as  
\begin{align}
&\bra \, Y \, M | \mathcal{L}_{tH} 
\, | \, N \, \pi^- \ket \cr
&\propto\,\delta^{(3)}\big(\vec{P}_Y -\vec{P}_N-\vec{q}\big)
\cr
&\hspace{0.3cm}\times \int d^3 q_2\,d^3 q_3 
\,\delta^{(3)}\big(\vec{q} -\vec{q}_2-\vec{q}_3\big)
\int d^3 \rho e^{i \vec{q}_\rho \cdot \vec{\rho}}
\psi^{\rho *}_{l_{\rho}}(\vec{\rho}) \psi^{\rho }_{0} (\vec{\rho})
\int d^3 \lambda e^{i \vec{q}_\lambda \cdot \vec{\lambda}}
\psi^{\lambda' *}_{l_{\lambda}}(\vec{\lambda}) \psi^{\lambda}_{0} (\vec{\lambda})
\nonumber \\
&\hspace{0.2cm}
 + (1 \leftrightarrow 2, \, \vec{\rho} \rightarrow -\vec{\rho}) 
\end{align}
Having performed the integration with respect to $q_2$, $q_3$, and
$\rho$, $\lambda$, we obtain the matrix elements for heavy baryon
productions as 
\begin{align}
\bra Y(l_\rho, l_\lambda) \hspace{0.1cm} M | \mathcal{L}_{tH} |
  p\hspace{0.1cm} \pi^- \ket &=C_{YM} \, I_{l} \,
                               (2\pi)^3\delta^{(3)}\big(\vec{P}_Y
                               -\vec{P}_p-\vec{q}\big) 
\label{eq:TransAmpGn}
\end{align}
where $I_{l}$ is given as 
\begin{align}
I_{0} &\equiv
\left(
\frac{16\pi \alpha^2_{\rho} \alpha_{\lambda'}\alpha_{\lambda}}{
	B^2}
\right)^{\frac{3}{2}}
e^{-q^2_{eff}/(4 B^2)},
\end{align}
for $l=0$
\begin{align}
I_{l_\lambda=1} &\equiv
\frac{i\sqrt{2}\alpha_{\lambda'} |\vec{q}_{eff}|}{
  2 B^2}
\left(
\frac{16\pi \alpha^2_{\rho} \alpha_{\lambda'}\alpha_{\lambda}}{
  B^2}
\right)^{\frac{3}{2}}
e^{-q^2_{eff}/(4 B^2)}
\end{align}
and
\begin{align}
I_{l_\rho=1} &\equiv 
\frac{-i\sqrt{2}\alpha_{\rho} |\vec{q}_{eff}|}{
  B^2}
\left(\frac{16\pi \alpha^2_{\rho} \alpha_{\lambda'}\alpha_{\lambda}}{
  B^2}\right)^{\frac{3}{2}}
e^{-q^2_{eff}/(4 B^2)}
\end{align}
for $l=1$. The coefficients $C_{YM}$ are the prefactors from spin and
isospin calculations. The effective momentum transfer, $\vec{q}_{eff}$ and
$B^2$ are defined as
\begin{align}
\vec{q}_{eff}
&\equiv\frac{m_d }{m_d + m_q}\vec{P}_N - \frac{m_d }{m_d + m_Q}\vec{P}_Y,
\hspace{0.2cm}
B^2 \equiv \frac{8 \alpha_\rho^2+\alpha_{\lambda'}^2+\alpha_{\lambda}^2}{2} 
\end{align}
where $m_d$, $m_q$, and $m_Q$ are the masses of a diquark, light
quarks ($u$, $d$ quark), and the heavy quark,
respectively. $\alpha_{\rho}$, $\alpha_{\lambda}$, and
$\alpha_{\lambda'}$ are the oscillator parameters for the
$\rho-$modes, initial and final state $\lambda-$modes, respectively.  
Here, except for the delta function, the matrix elements
Eq. (\ref{eq:TransAmpGn}) depend on $\vec{q}_{eff}$ instead of
$\vec{q}$ because of the so-called recoil effect due to the change in
the masses of particles before and after the interaction. 

In the center of mass frame, the differential cross sections for the
heavy baryon productions can be written as 
\begin{align}
\mathcal{R}\big(Y(J^{p}, J_z) \big)
=\frac{1}{4|p_{i}|\sqrt{s}} |C_{YM}|^2 |I_l|^2 \frac{|\vec{p}_f|}{4\pi \sqrt{s}}.
\label{eq:ProdRates}
\end{align}

\section{Discussion}
The production rates of hyperons and charmed baryons are given in
Table~\ref{tab:1}. In the present work, we calculate the production rates
of various heavy baryon productions. Since this work is the first
attempt of the two-quark process, we consider a simple case, i.e. the 
forward angle scattering. Those at finite angles is left for future
works. To demonstrate the production rates, we set 
the momentum of the pion at $k^{Lab}_\pi = 5\, \mathrm{GeV}$ for the
hyperons  and $k^{Lab}_\pi = 20\, \mathrm{GeV}$ for charmed
baryons. These momenta provide sufficient energies to create $s\bar s$
or $c \bar c$  pair.  In the two-quark process, the momentum $\vec{q}$
is shared by the heavy quark and a light quark in the heavy baryon,
which may excite both $\lambda$- and $\rho$-modes. 
This contrasts with the one-quark process where only one quark
receives the transferred momentum and possible excitations are only
$\lambda$-modes. 
 
A more detailed discussion about momentum-transfer and structure
dependences of the production rates will be given elsewhere.   

\begin{table}[!htbp]
\centering
\small
\caption{ Production rates of various heavy baryons, $\mathcal{R}(Y)$,
  which are normalized by that of the ground-state
  $\Lambda(\frac{1}{2}^+)$ where $Y_s$ and $Y_c$ are for the strange
  and charmed baryons, respectively. $j$ is a coupled spin, so-called
  a brown muck spin,  by spin and orbital angular momentum of a
  diquark in the baryons.} 
\begin{tabular}{lccccccc}\hline\hline \vspace{-0.3cm}\\
$l=0$    & $\Lambda\left(\frac{1}{2}^+\right)$
     & $\Sigma\left(\frac{1}{2}^+\right)$
     & $\Sigma\left(\frac{3}{2}^+\right)$ & & & & \\ 
$\mathcal{R}$($Y_s$) &  1   & 3.2  & 0    &   &    &  & \\
$\mathcal{R}$($Y_c$) &  1   & 2.9  & 0    &   &    &  & \\
\hline  \vspace{-0.3cm} \\
$l_\lambda=1$
& $\Lambda\left(\frac{1}{2}^-\right)$ 
& $\Lambda\left(\frac{3}{2}^-\right)$ 
& $\Sigma\left(\frac{1}{2}^-\right)$
& $\Sigma\left(\frac{1}{2}^-\right)$ 
& $\Sigma\left(\frac{3}{2}^-\right)$
& $\Sigma\left(\frac{3}{2}^-\right)$ 
& $\Sigma\left(\frac{5}{2}^-\right)$  \\
& $j=1$ & $j=1$ & $j=0$ & $j=1$ & $j=1$ & $j=2$ & $j=2$ \\
$\mathcal{R}$($Y_s$) & 0.004& 0.010& 0.007& 0.015& 0.007& 0.038 & 0\\
$\mathcal{R}$($Y_c$) & 0.10 & 0.20  & 0.12 & 0.23  & 0.12 & 0.58  & 0\\
\hline \vspace{-0.3cm} \\
$l_\rho=1$  
& $\Lambda\left(\frac{1}{2}^-\right)$ 
& $\Lambda\left(\frac{1}{2}^-\right)$ 
& $\Lambda\left(\frac{3}{2}^-\right)$
& $\Lambda\left(\frac{3}{2}^-\right)$ 
& $\Lambda\left(\frac{5}{2}^-\right)$
& $\Sigma \left(\frac{1}{2}^-\right)$ 
& $\Sigma \left(\frac{3}{2}^-\right)$ \\
& $j=0$ & $j=1$ & $j=1$ & $j=2$ & $j=2$ & $j=1$ & $j=1$ \\
$\mathcal{R}$($Y_s$) & 0.017& 0.039 & 0.018& 0.10 & 0   & 0.016& 0.032\\
$\mathcal{R}$($Y_c$) & 0.22 & 0.43  & 0.22 & 1.1  & 0   & 0.20 & 0.41  \\
\hline \hline
\end{tabular}
\label{tab:1}
\end{table}

\section*{Acknowledgments}
SIS and AH are supported in part by Grant-in-Aid for Scientific
Research on Innovative Areas, Clustering as a window on the
hierarchical structure of quantum systems. HChK is supported by the 
Basic Science Research Program through the National Research
Foundation (NRF) of Korea funded by the Korean 
government (MEST):
Grant No. 2018R1A5A1025563. 



\end{document}